\documentclass[intlimits,twoside,a4paper]{article}

\usepackage{amsmath,amssymb}
\usepackage{graphicx}
\usepackage{wrapfig}

\usepackage[T2A]{fontenc}
\usepackage[cp1251]{inputenc}

\usepackage{epsf,multicol,ifthen}
\usepackage[eqsecnum]{cmpj2}

\issue{2016}{19}{4}{43801}
\doinumber{10.5488/CMP.19.43801}
\title[Interaction between a surface acoustic wave and adsorbed atoms ]%
{Interaction between a surface acoustic wave and adsorbed atoms}
\author[R.M.~Peleshchak, M.Ya.~Seneta]{R.M.~Peleshchak, M.Ya.~Seneta\thanks{E-mail: marsen18@i.ua}
}
\address{Drohobych Ivan Franko State Pedagogical University, 24 Franko St., 82100 Drohobych, Ukraine}

\authorcopyright{R.M.~Peleshchak, M.Ya.~Seneta, 2016}
\date{Received October 4, 2016, in final form December 2, 2016}

\DeclareMathOperator{\grad}{grad}
\DeclareMathOperator{\diver}{div}

\begin{document}

\maketitle

\begin{abstract}
Within nonlocal elastic interaction between an adsorbed atom and matrix atoms and in consideration of the mirror image forces, a dispersion law of elastic surface acoustic waves is found to the long-wavelength approximation depending on the concentration of adsorbed atoms and the deformation potential of the adatom. The energy width of the surface acoustic mode is calculated depending on the concentration of adsorbed atoms.

\pacs 81.07.Bc, 66.30.Lw
\end{abstract}

\section{Introduction}
The research of electronic and phonon processes and their interaction on the defective semiconductor surface is indispensable for the development of state-of-the-art microelectronic and nanoelectronic technologies. In particular, the mechanisms of interaction between adsorbed atoms and the surface acoustic wave (SAW) caused by the deformation potential and its effect on the formation of surface electronic states, dispersion and the surface acoustic wave damping is an issue of great importance. As it was stated in~\cite{Vla15}, surface acoustic waves can be a source of long-range effects that induce the formation of nanoclusters on a crystal surface beyond the laser irradiation area. Therefore, the research of the acoustic wave damping processes on the monocrystalline substrate defective surface is urgent for the development of optimal technological conditions of nanostructure formation at the nanosecond laser irradiation of the CdTe surface \cite{Vla15} as well as for the construction of nanoelectronic devices.

The theory of the initial stage of formation (nucleation) of periodic nanometer adatoms structure due to the instability caused by the interaction between the adatoms and the self-consistent static ($\omega =0$) acoustic quasi-Rayleigh SAW~\cite{Lan70} with exponentially increasing amplitude was developed in~\cite{Eme02}.

Moreover, the surface acoustic wave (SAW) method is widely used to diagnose the dynamic properties of two-dimensional electronic layers (e.g., the dynamic conductivity, the carrier mobility or carrier concentration) of GaAs/In$_{1-x}$Ga$_{x}$As/GaAs, Al$_{x}$Ga$_{1-x}$As/GaAs/Al$_{x}$Ga$_{1-x}$As, Cd$_{1-x}$Zn$_{x}$Te/CdTe/Cd$_{1-x}$Zn$_{x}$Te nanoheterostructures \cite{Fil99,Wix86} due to their inhomogeneous strain and piezoelectric fields. The surface acoustic wave (SAW) generates a variable electric field that interacts with two-dimensional electrons. This leads to renormalization of both the velocity and the SAW damping coefficient.

The purpose of this paper is to study the effect of interaction between adatoms and the self-consistent acoustic quasi-Rayleigh wave on its dispersion and phonon mode width at different concentrations of adsorbed atoms.

\section{The model of deformation interaction between adatoms and the surface acoustic wave (SAW)}

\begin{figure}[!b]
   \begin{center}
   \includegraphics[width=300pt]{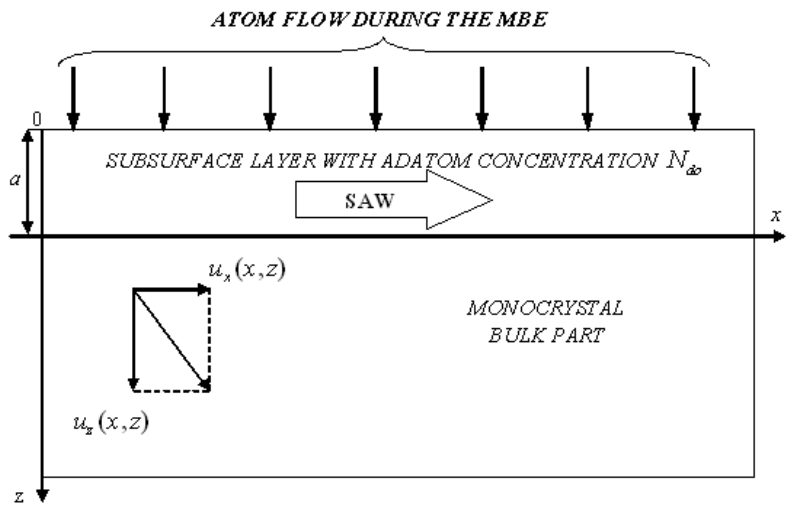}
   \caption{Interaction of the SAW with the subsurface adatom layer.}
   \label{fig1}
   \end{center}
\end{figure}

Figure \ref{fig1} shows a structure consisting of the semiconductor substrate where $a$ is the subsurface defective layer thickness.

The atoms adsorbed during the molecular beam epitaxy or implantation could be considered as surface defects. The subsurface layer is non-uniformly deformed by adatoms due to the deformation potential and the local surface energy renormalization. This inhomogeneous self-consistent deformation, in its turn, redistributes the adsorbed atoms across the surface through the deformation potential. Therefore, the effect of the adsorbed atoms reduces to the change of boundary conditions for the stress tensor $\sigma _{ij} $ on the $z=0$ surface.

The displacement vector $\textbf{u}(\textbf{r},t)$ of the points of the medium satisfies the following equation~\cite{Lan70}:

\begin{equation}
\label{eq1}
\frac{\partial ^{2} \textbf{u}}{\partial t^{2} } =c_{t}^{2} \Delta _{\textbf{r}} \textbf{u}+\left(c_{l}^{2} -c_{t}^{2} \right)\grad(\diver\textbf{u}).
\end{equation}

In the geometrical model (figure~\ref{fig1}), the solution of equation (\ref{eq1}) for the surface Rayleigh wave spreading along the $x$ axis is:

\begin{equation}
\label{eq2}
u_{x} (x,z)=-\ri qA\re^{\ri qx-\ri \omega t-k_{l} z} -\ri k_{t} B\re^{\ri qx-\ri\omega t-k_{t} z},
\end{equation}

\begin{equation}
\label{eq3}
u_{z} (x,z)=k_{l} A\re^{\ri qx-\ri \omega t-k_{l} z} +qB\re^{\ri qx-\ri\omega t-k_{t} z},
\end{equation}
where $k_{l,t}^{2} =q^{2} -\omega ^{2} /c_{l,t}^{2}  $ and $A$, $B$ are the SAW amplitudes.

The $x$ direction on the crystal surface is determined due to the elastic anisotropy, while on the isotro\-pic surface it could be determined due to an external effect, which induces the elastic anisotropy, or due to spontaneous symmetry breaking of the defect-deformation system, similarly to ~\cite{Hak80}.

The deformation $\varepsilon $ on the surface of the semiconductor ($z$=0) is bound with the components of the displacement vector by the relation

\begin{equation}
\label{eq4}
\varepsilon (x,t)=\frac{\partial u_{x} }{\partial x} +\frac{\partial u_{z} }{\partial z}=\frac{\omega ^{2} }{c_{l}^{2} } A\re^{\ri qx-\ri\omega t}.
\end{equation}

Spatially inhomogeneous surface deformation $\varepsilon (x,t)$ leads to the inhomogeneous adatom redistribution $N_{\text{d}} (x,t)$.

\begin{equation}
\label{eq5}
N_{\text{d}} (x,t)=N_{0\text{d}} +N_{1\text{d}} (x,t)=N_{0\text{d}} +N_{1\text{d}} (q)\re^{\ri qx-\ri\omega t},
\end{equation}
where $N_{0\text{d}} $ is the spatially homogeneous component; $N_{1\text{d}} (q)$ is the periodic disturbance amplitude \linebreak $[N_{1\text{d}}(x)\ll N_{0\text{d}}]$.

Now that we have the deformation parameter $\varepsilon (x,t)$, we can find the interaction energy between the adsorbed atom and the matrix atoms $W_{\text{da}} $ through the elastic field~\cite{Eme97,Pel14} as an approximation of nonlocal Hooke law~\cite{Kun70}:

\begin{equation}
\label{eq6}
W_{\text{da}} (x)=-\int \lambda (|x'-x|)\varepsilon (x')\Delta\Omega _{\text{d}} \rd x',
\end{equation}
where $\lambda $ is the elastic moduli operator [9]; $\Delta\Omega _{\text{d}} $ is the crystal volume change caused by a single adsorbed atom.
Introducing the variable $\tau =x'-x$ and expanding into a Taylor series, we obtain:
\begin{align}
\label{eq7}
W_{\text{da}}^{\text{int}} (x)&=-\int \lambda \left(\left|\tau \right|\right)\varepsilon (x+\tau )\Delta \Omega _{\text{d}} \rd\tau  =-\int \lambda \left(\left|\tau \right|\right)\left[\varepsilon (x)+\frac{\partial ^{2} \varepsilon (x)}{\partial x^{2} } \frac{\tau ^{2} }{2} \right]\Delta \Omega _{\text{d}} \rd\tau  \nonumber\\ &=-K_{\text{d}} \varepsilon (x)\Delta \Omega _{\text{d}} -K_{\text{d}} \frac{\partial ^{2} \varepsilon (x)}{\partial x^{2} } l_{\text{d}}^{2} \Delta \Omega _{\text{d}}\, ,
\end{align}
where $K_{\text{d}} =\int \lambda(|\tau|) \rd\tau \equiv K$ is the modulus of elasticity \cite{Eme97};
$l_{\text{d}}^{2} =\int \lambda (\tau )\tau ^{2} \rd\tau  /[2\int \lambda (|\tau |) \rd\tau ] $ is the average of the square of characteristic distance of interaction between the adatom and the matrix atoms.

The elastic fields of adsorbed atoms shift the atoms into the neighborhood of other adatoms and create the forces affecting them. This is caused by the elastic interaction between them. The energy of this interaction decreases in accord with the power law and is rather significant when the crystal lattice is heavily deformed by the adatoms \cite{Kun70}. Within isotropic solids, the elastic defect interaction energy is equal to zero~\cite{Kri83}.

The elastic interaction of the adsorbed atoms decreases according to the power law as the distance increases. Besides, there is another interaction that gradually changes at distances comparable to the crystal size and is connected with mirror image forces applied to the crystal surface \cite{Kri83}, which provide boundary conditions on the crystal surface (e.g., the condition with no stresses on the crystal surface). The field of such forces is called the image field or the imaginary sources field similarly to the electrostatic field of charge arising on the conductive surface and is equivalent to the mirror image charge field \cite{Kos88}. The interaction energy $W_{\text{da}}^{\text{int}} $ between an adsorbed atom in the position $r'$ and other adatoms having concentration $N_{\text{d}} (x)$ caused by these forces, is virtually independent of the adatom position $r'$ and can be defined as in~\cite{Kri83}:

\begin{equation}
\label{eq8}
W_{\text{dd}}^{\text{int}} (x)=-\frac{2}{3} \frac{1-2\nu }{K(1-\nu )a} \theta _{\text{s}}^{2} N_{\text{d}} (x),
\end{equation}
where $\nu $ is the Poisson coefficient;
$\theta _{\text{s}}^{} =K\cdot \Delta \Omega _{\text{d}} $ is the surface deformation potential.

The force \textit{F} acting on the adatom due to the elastic field appearing in the implanted adatoms matrix is given by:

\begin{equation}
\label{eq9}
F=-\frac{\partial \left[W_{\text{dd}}^{\text{int}} (x)+W_{\text{da}}^{\text{int}} (x)\right]}{\partial x}=\frac{2}{3} \frac{1-2\nu }{K(1-\nu )a} \theta _{\text{s}}^{2} \frac{\partial N_{\text{d}} (x)}{\partial x} +\theta _{\text{s}} \frac{\partial \varepsilon (x,t)}{\partial x} +\theta _{\text{s}} l_{\text{da}}^{2} \frac{\partial ^{3} \varepsilon (x,t)}{\partial x^{3} } \,,
\end{equation}
which induces the regular diffusion flow $\left[-D_{\text{d}} \frac{\partial N_{\text{d}} (x)}{\partial x} \right]$ and the additional deformation flow of adatoms. The latter is due to the deformation gradients $\frac{\partial \varepsilon (x,t)}{\partial x} ,{\rm \; }\frac{\partial }{\partial x} \left[\frac{\partial ^{2} \varepsilon (x,t)}{\partial x^{2} } \right]$, the defect concentration $\frac{\partial N_{\text{d}} (x)}{\partial x} $, and the monocrystal subsurface layer volume $\Delta \Omega _{\text{d}} $ change caused by these adatoms.

The analysis of formula (\ref{eq9}) shows that the concentration gradient creates a deformation flow component which is directed towards the side where the adatom concentration is greater (the first term), unlike a regular diffusion flow. Furthermore, the adatoms being the extension centers ($\Delta \Omega _{\text{d}} >0$) will move to the area that experience a tensile strain, while the adatoms being the compression centers ($\Delta \Omega _{\text{d}} <0$) will move to the area of a relative compression (the second term).

Under this force (\ref{eq9}), the adatoms in the elastic field get the velocity

\begin{equation}
\label{eq10}
\upsilon =\mu  F=\frac{2}{3} \frac{1-2\nu }{K(1-\nu )} \frac{D_{\text{d}} \theta _{\text{s}}^{2} }{k_{\text{B}} Ta} \frac{\partial N_{\text{d}} (x)}{\partial x} +\frac{D_{\text{d}} \theta _{\text{s}}^{} }{k_{\text{B}} T} \frac{\partial \varepsilon (x,t)}{\partial x} +\frac{D_{\text{d}} \theta _{\text{s}}^{} }{k_{\text{B}} T} l_{\text{da}}^{2} \frac{\partial ^{3} \varepsilon (x,t)}{\partial x^{3} } \,,
\end{equation}
where $D_{\text{d}} $ is the adatom diffusion coefficient; $T$ is temperature; $k_{\text{B}} $ is the Boltzmann constant. Here, the Einstein relation is used to determine the mobility $\mu $ of adatoms.

Taking into account \eqref{eq10} and the continuity equation $\diver\textbf{j}=-\frac{\partial N_{\text{d}} (x,t)}{\partial t} $, the flow of the implanted adatoms and the equation for the concentration of adatoms can be written as follows:

\begin{equation}
\label{eq11}
j=-D_{\text{d}} \frac{\partial N_{\text{d}} (x,t)}{\partial x}+\frac{D_{\text{d}} \theta _{\text{s}}^{} }{k_{\text{B}} T} N_{\text{d}} (x,t)\frac{\partial }{\partial x} \left[\frac{2}{3} \frac{1-2\nu }{K(1-\nu )a} \theta _{\text{s}} N_{\text{d}} (x,t)+\varepsilon (x,t)+l_{\text{da}}^{2} \frac{\partial ^{2} \varepsilon (x,t)}{\partial x^{2} } \right],
\end{equation}

\begin{align}
\label{eq12}
\frac{\partial N_{\text{d}} (x,t)}{\partial t} &=D_{\text{d}} \frac{\partial ^{2} N_{\text{d}} (x,t)}{\partial x^{2} }-\frac{D_{\text{d}} \theta _{\text{s}} }{k_{\text{B}} T} \frac{\partial }{\partial x} \left\{N_{\text{d}} (x,t)\frac{\partial }{\partial x} \left[\frac{2}{3} \frac{1-2\nu }{K(1-\nu )a} \theta _{\text{s}} N_{\text{d}} (x,t)\right.\right.\nonumber\\&\quad\left.\left.+\varepsilon (x,t)+l_{\text{da}}^{2} \frac{\partial ^{2} \varepsilon (x,t)}{\partial x^{2} } \right]\right\}.
\end{align}

The first term in (\ref{eq12}) describes the regular gradient concentration diffusion, while the second one describes a qualitatively new diffusion effect of the ``flow of deformation retraction'' caused by mirror image forces and the deformation gradient~\cite{Fal95} as well as the nonlocal interaction between the adatoms and the surface atoms~\cite{Pel14}.

Considering the condition $N_{\text{d1}}\ll N_{\text{d0}} $ and (\ref{eq5}), the equation (\ref{eq12}) to the linear approximation will be:

\begin{equation}
\label{eq13}
\left\{-\ri\omega +D_{\text{d}} \left[1-\frac{2}{3} \frac{1-2\nu }{K(1-\nu )a} \frac{\theta _{\text{d}} ^{2} }{k_{\text B}T} N_{\text{d0}} \right]q^{2} \right\}N_{\text{1d}} (q)=\frac{D_{\text{d}} \theta _{\text{d}} }{k_{\text B}T} N_{\text{d0}} \varepsilon (q)q^{2} \left(1-l_{\text{d}}^{2} q^{2} \right).
\end{equation}
From the equation (\ref{eq13}) we obtain the expression for the amplitude of the surface adatom concentration~$N_{\text{d1}}(q)$.

Spatially inhomogeneous adatom distribution modulates the surface energy

\[F(x)=F_{0} +\frac{\partial F}{\partial N_{\text{d1}} } N_{\text{d1}} {\rm (}x{\rm )},\]
which results in the lateral mechanical stress

\[\sigma _{xz} =\frac{\partial F\big(N(x)\big)}{\partial x} \,,\]
that is compensated by the displacement stress within the medium~\cite{Eme02}. The boundary condition expressing the balance of lateral stresses is as follows:

\begin{equation}
\label{eq14}
\left. \mu \left(\frac{\partial u_{x} }{\partial z} +\frac{\partial u_{z} }{\partial x} \right)\right|_{z=0} =\frac{\partial F}{\partial N_{\text{d1}} } \frac{\partial N_{\text{d1}} (x)}{\partial x}\,,
\end{equation}
where $\mu $ is the displacement module of the medium.

Herein below we are considering the coefficient $\frac{\partial F}{\partial N_{\text{d1}}} $ as the predetermined phenomenological parameter.

In addition, the interaction between the adatoms and the semiconductor atoms results in a normal mechanical stress on the surface. The corresponding boundary condition is as follows:

\begin{equation}
\label{eq15}
\left. \left[\frac{\partial u_{z} }{\partial z} +(1-2\beta )\frac{\partial u_{x} }{\partial x} \right]\right|_{z=0} =\frac{\theta _{\text{d}} N_{\text{d1}} (x)}{\rho c_{l}^{2} a}\,,
\end{equation}
where $a$ is the crystal lattice parameter on the semiconductor surface; $\beta =c_{t}^{2} /c_{l}^{2} $; $\rho $ it is the crystal density.

\section{Dispersion equation of the surface acoustic wave interacting with adsorbed atoms}

To obtain the dispersion equation, we substitute (\ref{eq13}) into (\ref{eq14}) and (\ref{eq15}), taking into account (\ref{eq4}) and (\ref{eq5}) as well as (\ref{eq2}) and (\ref{eq3}). As a result, we obtain a system of two linear equations for the amplitudes $A$ and $B$. Then, from the condition of nontrivial solutions, we obtain the dispersion equation for the surface acoustic wave that interacts with the adsorbed atoms:
\begin{align}
\label{eq16}
\left(q^{2} +k_{t}^{2} \right)^{2} -4q^{2} k_{l} k_{t} &=-\frac{2}{\beta } \frac{\omega ^{2} }{c_{l}^{2} } \frac{\theta _{\text{d}} N_{\text{d0}} }{k_{\text B}T\rho c_{l}^{2} }\frac{D_{\text{d}} q^{2} }{-\ri\omega +D_{\text{d}} \Big[1-\frac{2}{3} \frac{1-2\nu }{K(1-\nu )a} \frac{\theta _{\text{d}} ^{2} }{k_{\text B}T} N_{\text{d0}}\Big]q^{2} }\left(1-l_{\text{d}}^{2} q^{2} \right) \nonumber\\ &\quad\times \left[q^{2} k_{t} \frac{\partial F}{\partial N_{\text{d1}} } +\left(q^{2} +k_{t}^{2} \right)\frac{\theta _{\text{d}} }{2a} \right].
\end{align}

The left-hand part of (\ref{eq16}) matches the Rayleigh determinant; the latter, when set equal to zero, determines the dispersion law of the Rayleigh surface acoustic wave without adsorbed atoms \cite{Lan70}. The right-hand part of (\ref{eq16}) renormalizes the dispersion equation of the Rayleigh acoustic wave \cite{Lan70} due to the adsorbed atoms power action ($\sim \theta _{\text{d}} $) causing the strain of the crystal lattice subsurface layer. Substituting $\omega =c_{t} q\xi $ into (\ref{eq16}), we obtain:
\begin{align}
\label{eq17}
\left(2-\xi ^{2} \right)^{2} -4\left(1-\xi^{2}\right)^{\frac{1}{2}}\left(1-\frac{c_{t}^{2} }{c_{l}^{2} } \xi^{2} \right)^{\frac{1}{2}}&=-\frac{2\xi ^{2} \theta _{\text{d}} N_{\text{d0}} }{k_{\text B}T\rho c_{l}^{2} } \frac{D_{\text{d}} q\left\{D_{\text{d}} \Big[1-\frac{2}{3} \frac{1-2\nu }{K(1-\nu )a} \frac{\theta _{\text{d}} ^{2} }{k_{\text B}T} N_{\text{d0}} \Big]q+\ri c_{t}\xi\right\}}{\left\{D_{\text{d}} \Big[1-\frac{2}{3} \frac{1-2\nu }{K(1-\nu )a} \frac{\theta _{\text{d}} ^{2} }{k_{\text B}T} N_{\text{d0}} \Big]\right\}^{2} q^{2} +c_{t}^{2} \xi ^{2} } \nonumber\\ &\quad\times\left(1-l_{\text{d}}^{2} q^{2} \right)\left[q\left(1-\xi ^{2}\right)^{\frac{1}{2}} \frac{\partial F}{\partial N_{\text{d1}} } +\left(2-\xi ^{2} \right)\frac{\theta _{\text{d}} }{2a} \right].
\end{align}

Expression (\ref{eq17}) has a real part and an imaginary part that finally determine a correction to the dispersion law of Rayleigh wave and its damping. The multiplier $q$ in the numerator of (\ref{eq17}) allows us to solve this equation using the iterative method in the long-wave region $qa\ll1$.

Let the left-hand part of (\ref{eq17}) be denoted by the function $f(\xi )$ and let us expand the latter into a Taylor series in the vicinity of the point $\xi_{0}$
\begin{equation}
\label{eq18}
f(\xi _{0} +\delta \xi )\approx f(\xi _{0} )+f'(\xi _{0} )\delta \xi,
\end{equation}
where $\xi _{0} $ is the solution of the equation $f(\xi _{0} )=0$.

Then, the correction $\delta \xi $ is determined by the right-hand part of (\ref{eq17}) with the substitution $\xi \rightarrow \xi _{0} $.
\begin{align}
\label{eq19}
\delta \xi &=-\frac{1}{f'(\xi _{0} )} \frac{2\xi _{0} ^{2} \theta _{\text{d}} N_{\text{d0}} }{k_{\text B}T\rho c_{l}^{2} } \frac{D_{\text{d}} ^{2} q^{2} \Big[1-\frac{2}{3} \frac{1-2\nu }{K(1-\nu )a} \frac{\theta _{\text{d}} ^{2} }{k_{\text B}T} N_{\text{d0}} \Big]}{\left\{D_{\text{d}} \Big[1-\frac{2}{3} \frac{1-2\nu }{K(1-\nu )a} \frac{\theta _{\text{d}} ^{2} }{k_{\text B}T} N_{\text{d0}} \Big]\right\}^{2} q^{2} +c_{t}^{2} \xi _{0} ^{2} }
\left(1-l_{\text{d}}^{2} q^{2} \right)\left[q\left(1-\xi _{0} ^{2} \right)^{\frac{1}{2}} \frac{\partial F}{\partial N_{\text{d1}} } \right.
\nonumber \\ &\left.\quad+\left(2-\xi _{0} ^{2} \right)\frac{\theta _{\text{d}} }{2a} \right]
 -\ri\frac{1}{f'(\xi _{0} )} \frac{2\xi _{0} ^{3} c_{t} \theta _{\text{d}} N_{\text{d0}} }{k_{\text B}T\rho c_{l}^{2} } \frac{D_{\text{d}} q}{\left\{D_{\text{d}} \Big[1-\frac{2}{3} \frac{1-2\nu }{K(1-\nu )a} \frac{\theta _{\text{d}} ^{2} }{k_{\text B}T} N_{\text{d0}} \Big]\right\}^{2} q^{2} +c_{t}^{2} \xi _{0} ^{2} } \left(1-l_{\text{d}}^{2} q^{2} \right)
\nonumber\\ &\quad\times\left[q\left(1-\xi _{0} ^{2} \right)^{\frac{1}{2}} \frac{\partial F}{\partial N_{\text{d1}} } +\left(2-\xi _{0} ^{2} \right)\frac{\theta _{\text{d}} }{2a} \right].
\end{align}

The numerical analysis shows that $f'(\xi _{0} )>0$ in the whole region of change of $\xi _{0} $.

Separating the real and imaginary parts in (\ref{eq19}) and considering $\omega =c_{t} q\xi _{0} +c_{t} q\delta \xi $, we obtain the expressions for the dispersion law $\omega '(q)$ of the elastic surface acoustic wave and its width. The latter is caused by the interaction between the adsorbed atoms and the self-consistent quasi-Rayleigh wave in consideration of both the nonlocal elastic interaction of implanted impurity with the matrix atoms~\cite{Pel14}, and the mirror image forces~\cite{Kri83}:
\begin{align}
\label{eq20}
&\omega '(q)=c_{t} q\xi _{0} \left(1-\frac{1}{f'(\xi _{0} )} \frac{2\xi _{0} \theta _{\text{d}} N_{\text{d0}} }{k_{\text B}T\rho c_{l}^{2} } \frac{D_{\text{d}} ^{2} q^{2} \Big[1-\frac{2}{3} \frac{1-2\nu }{K(1-\nu )a} \frac{\theta _{\text{d}} ^{2} }{k_{\text B}T} N_{\text{d0}} \Big]}{\left\{D_{\text{d}} \Big[1-\frac{2}{3} \frac{1-2\nu }{K(1-\nu )a} \frac{\theta _{\text{d}} ^{2} }{k_{\text B}T} N_{\text{d0}} \Big]\right\}^{2} q^{2} +c_{t}^{2} \xi _{0} ^{2} }\left(1-l_{\text{d}}^{2} q^{2} \right)
\right.
\nonumber\\ &\quad\,\,\left.\phantom{\frac{\frac{x^{2}}{\frac{x^{\frac{1}{2}}}{x^{2}}}}{\frac{x^{2}}{x^{2}}}}\times
\left[q\left(1-\xi _{0} ^{2} \right)^{\frac{1}{2}} \frac{\partial F}{\partial N_{\text{d1}} } +\left(2-\xi _{0} ^{2} \right)\frac{\theta _{\text{d}} }{2a} \right]\right),\\
&\omega ''(q)=-c_{t}^{2} \frac{1}{f'(\xi _{0} )} \frac{2\xi _{0}^{3} \theta _{\text{d}} N_{\text{d0}} }{k_{\text B}T\rho c_{l}^{2} } \frac{D_{\text{d}} q^{2} }{\left\{D_{\text{d}} \Big[1-\frac{2}{3} \frac{1-2\nu }{K(1-\nu )a} \frac{\theta _{\text{d}} ^{2} }{k_{\text B}T} N_{\text{d0}} \Big]\right\}^{2} q^{2} +c_{t}^{2} \xi _{0} ^{2} } \left(1-l_{\text{d}}^{2} q^{2} \right)
\nonumber\\ &\,\,\,\,\,\quad\qquad\times\left[q\left(1-\xi _{0} ^{2} \right)^{\frac{1}{2}} \frac{\partial F}{\partial N_{\text{d1}} } +\left(2-\xi _{0} ^{2} \right)\frac{\theta _{\text{d}} }{2a} \right].
\end{align}

\begin{figure}[!b]
  \begin{center}
  \includegraphics[width=370pt]{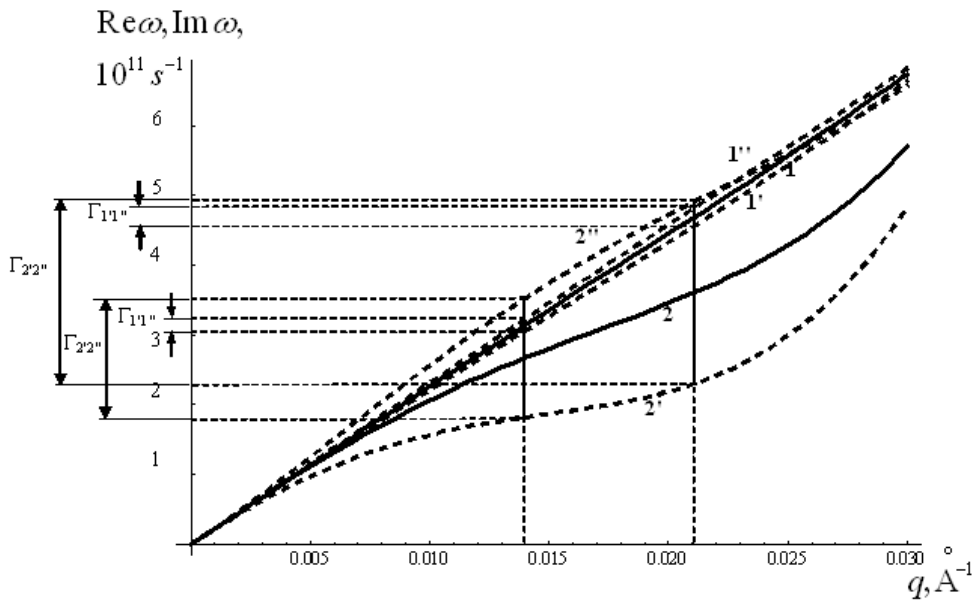}
  \caption{The dispersion law $\omega '(q)$ (curves $1$, $2$) and the phonon mode width $\omega ''(q)$ (curves $1'1''$, $2'2''$) of the surface acoustic wave interacting with the adsorbed atoms at the values of adsorbed atoms concentration $N_{\text{d0}} =3\cdot 10^{12}$~cm$^{-2}$ (curves $1$, $1'1''$); $3\cdot 10^{13}$~cm$^{-2}$ (curves $2$, $2'2''$) without considering the mirror image forces.}
  \label{fig2}
  \end{center}
\end{figure}

The numerical calculation of the elastic SAW dispersion law dependence $\omega '(q)=\Re\omega (q)$ and the acoustic mode width $\omega ''(q)=\Im\omega(q)$ has been conducted for the GaAs semiconductor with the adsorbed atom surface concentration $N_{\text{d0}} =3\cdot 10^{12}$~cm$^{-2}$ and $3\cdot 10^{13}$~cm$^{-2}$ at the following parameter values:
$l_{\text{d}} =2.9$~nm; $a=0.565$~nm; $c_{l} =4400$~m/s; $c_{t} =2475$~m/s; $\rho =5320$~kg/m$^3$; $D_{\text{d}} =5\cdot 10^{-2}$~cm$^2$/s; $\theta _{\text{d}} =10$~eV; $\frac{\partial F}{\partial N_{\text{d1}} } =0.1$~eV; $T=100$~K~\cite{Eme02}.

The characteristic length value $l_{\text{d}} $ of interaction between the adatom and the lattice atoms has been found from the condition of free energy minimum of the crystal with adsorbed atoms~\cite{Kun70}.

Figure~\ref{fig2} shows the results of calculation of the elastic surface acoustic mode frequency $\omega'(q)$ dependence (figure~\ref{fig2}, curves~$1$, $2$) on the module of wave vector $q$ and its width $\omega ''(q)$ (figure~\ref{fig2}, curves~$1'1''$, $2'2''$) caused by the interaction between the adsorbed atoms and the self-consistent acoustic quasi-Rayleigh wave at two values of the adsorbed atoms concentration $N_{\text{d0}} $ without considering the mirror image forces.

The dependence $\omega '(q)$ in the range of the wave vector module change $0\leqslant q<1/l_{\text{d}}  $ is nonlinear, while the dependence $\omega ''(q)$ is nonmonotonous.

At $q\rightarrow0 $, the surface acoustic mode width $\omega ''(q)$ tends to zero, while the dispersion curve $\omega '(q)$ asymptotically approaches the surface Rayleigh wave dispersion curve according to the dispersion law $\omega (q)=c_{t} \xi _{0} q$. It should be noted that at $q=1/l_{\text{d}}$, the surface acoustic wave-length is the same as the characteristic length of the adatom interaction with the lattice atoms. Figure~\ref{fig2} shows that the acoustic phonon mode width increases as the adsorbed atoms concentration increases. In particular, at $q=0.0014$~nm$^{-1} $ and the adatoms concentrations $N_{\text{d0}} =3\cdot 10^{12}$~cm$^{-2} $, the energy width $\Gamma_{1'1''}=\hbar \omega ''(q) $ of the surface acoustic mode is equal to $11.4$~\textmu eV, and at $N_{\text{d0}} =3\cdot 10^{13}$~cm$^{-2} $, the energy width $\Gamma_{2'2''}$ is equal to $114$~\textmu eV, while in more short-wave region of elastic surface acoustic waves ($q=0.0021$~nm$^{-1}$) at concentrations $N_{\text{d0}} =3\cdot 10^{12}$~cm$^{-2}$ and $3\cdot 10^{13}$~cm$^{-2} $, the energy width is equal to $17$~\textmu eV, $170$~\textmu eV, respectively.

\begin{figure}[!b]
  \begin{center}
  \includegraphics[width=370pt]{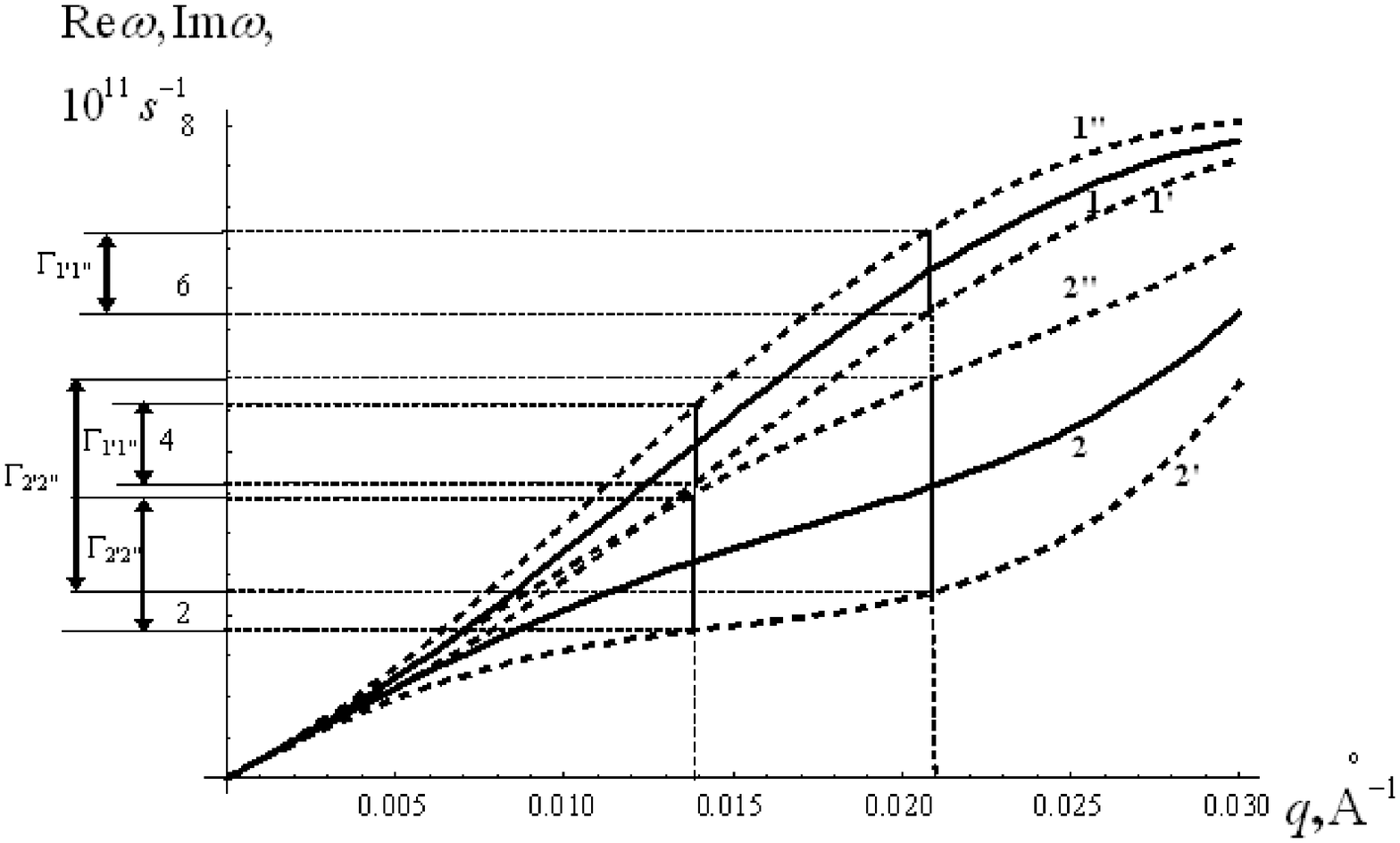}
  \caption{The dispersion dependence $\omega '(q)$ (curves $1$, $2$) and the elastic surface phonon mode width $\omega ''(q)$ (curves $1'1''$; $2'2''$)  interacting with adsorbed atoms with (curves $1$, $1'1''$) and without (curves $2$, $2'2''$) considering the mirror image forces at the value of adsorbed atoms concentration $N_{\text{d0}} =3\cdot 10^{13}$~cm$^{-2}$.}
  \label{fig3}
  \end{center}
\end{figure}

Figure~\ref{fig3} shows the dispersion dependences $\omega '(q)$ and the energy widths $\omega ''(q)$ of the surface acoustic modes at the adatom concentration $N_{\text{d0}} =3\cdot 10^{13}$~cm$^{-2} $ with (figure~\ref{fig3}, curves~$1$, $1'1''$) and without (figure~\ref{fig3}, curves~$2$, $2'2''$) considering the mirror image forces. In particular, at $q=0.0014$~nm$^{-1}$, $0.0021$~nm$^{-1}$, the energy widths of surface acoustic modes considering the mirror image forces are $63$~\textmu eV and $66$~\textmu eV, respectively, whereas without considering the mirror image forces, they are $114$~\textmu eV and $170$~\textmu eV, respectively. From the comparative analysis of the energy widths of acoustic modes we can see that the mirror image forces at $q=0.0014$~nm$^{-1}$ reduce the surface acoustic mode energy width by $51$~\textmu eV, while at $0.0021$~nm$^{-1} $, the reduction is by $104$~\textmu eV; thus, in the acoustic mode short-wave region, the effect of the mirror image forces on the change of the surface acoustic mode energy width is more important.

The character of dispersion law $\omega'(q)$, found in the long-wavelength limit ($qa\ll 1$, i.e., a neighbourhood of $\bar{\Gamma }$ of the surface BZ) and shown in figure~\ref{fig3} (curves~$1$, $2$), qualitatively converges with the dispersion law of the surface phonons in the $\bar{\Gamma }\bar{{\rm X}'}$ direction experimentally researched for the GaAs(110) surface without adsorbed atoms~\cite{Nie94}. The difference of the character of the dispersion law $\omega '(q)$ (figure~\ref{fig3}, curve~$2$) is observed at higher values of $q$ in the long-wavelength approximation. Such difference can be explained  by the fact that at a calculation of $\omega '(q)$ of curve $2$ (figure~\ref{fig3}), the mirror image forces were not taken into account, while at a calculation of $\omega '(q)$ of curve~$1$ (figure~\ref{fig3}), such forces were taken into account.

Practically, the surface acoustic waves (SAW) can be used as optical hologram reading transmitters in photorefractive crystals~\cite{Dee90}. The experimental data obtained by measuring the Raman spectra energy position change due to the nonelastic light scattering (Mandelstam-Brillouin), depending on the adsorbed atoms concentration and the strain potential, and can be used in the construction of selective gas sensors on the elastic surface acoustic modes~\cite{Wag89,Hup99}.

\section{Conclusions}

The elastic surface acoustic wave dispersion theory depending on the adsorbed atom concentration is developed within the nonlocal elastic interaction between  adsorbed atoms and matrix atoms in consideration of the mirror image forces.
We have established that:
\begin{enumerate}
\renewcommand{\labelenumi}{(\arabic{enumi})}
\item  the surface acoustic mode energy width is proportional to the product of the concentration of adsorbed atoms and the adsorbed atom surface deformation potential;
\item  the mirror image forces reduce the elastic surface acoustic mode energy width, while in the short-wave region, the effect of the mirror image forces on the change of the surface acoustic mode energy width is more important;
\item  at concentrations of adsorbed atoms located in the subsurface crystal lattice being the same as the interstitial impurities, the surface acoustic mode width is greater than in the case of the adsorbed atoms being substitutional impurities. This is because the surface deformation potential of the adsorbed atoms of the interstitial impurity type is greater than the surface strain potential of the adsorbed atoms of substitutional impurity type.
\end{enumerate}

\newpage
\ukrainianpart

\title{Взаємодія поверхневої акустичної хвилі з адсорбованими атомами}
\author[Р.М.~Пелещак, М.Я.~Сенета]{Р.М.~Пелещак, М.Я.~Сенета}
\address{Дрогобицький державний педагогічний університет імені Івана Франка, \\ вул. Івана Франка, 24, 82100 Дрогобич, Україна
}

\makeukrtitle

\begin{abstract}
\tolerance=3000%
У межах нелокальної пружної взаємодії адсорбованого атома з атомами матриці з врахуванням сил дзеркального зображення знайдено в довгохвильовому наближенні закон дисперсії поверхневих пружних акустичних хвиль в залежності від концентрації адсорбованих атомів і деформаційного потенціалу ад\-атома. Розраховано енергетичну  ширину поверхневої акустичної моди в залежності від концентрації адсорбованих атомів.

\keywords деформаційний потенціал, адатоми, сили дзеркального зображення, нелокальна пружна взаємодія, ширина поверхневої акустичної моди

\end{abstract}

\end{document}